\documentclass[11pt]{article}

\usepackage{hyperref}
\usepackage{graphicx}
\usepackage[square,comma]{natbib}
\setcitestyle{numbers}
\usepackage[margin=1in]{geometry}

\begin{document}

\title{The Hidden Assumptions Behind Counterfactual Explanations and Principal Reasons}

\author{Solon Barocas\thanks{Microsoft Research and Cornell University} \and Andrew D. Selbst\thanks{University of California Los Angeles} \and Manish Raghavan\thanks{ Cornell University}} 
\date{}

\maketitle
\begin{abstract}
Counterfactual explanations are gaining prominence within technical, legal, and business circles as a way to explain the decisions of a machine learning model. These explanations share a trait with the long-established ``principal reason'' explanations required by U.S. credit laws: they both explain a decision by highlighting a set of features deemed most relevant---and withholding others.

These ``feature-highlighting explanations'' have several desirable properties: They place no constraints on model complexity, do not require model disclosure, detail what needed to be different to achieve a different decision, and seem to automate compliance with the law. But they are far more complex and subjective than they appear.

In this paper, we demonstrate that the utility of feature-highlighting explanations relies on a number of easily overlooked assumptions: that the recommended change in feature values clearly maps to real-world actions, that features can be made commensurate by looking only at the distribution of the training data,  that features are only relevant to the decision at hand, and that the underlying model is stable over time, monotonic, and limited to binary outcomes.

We then explore several consequences of acknowledging and attempting to address these assumptions, including a paradox in the way that feature-highlighting explanations aim to respect autonomy, the unchecked power that feature-highlighting explanations grant decision makers, and a tension between making these explanations useful and the need to keep the model hidden.

While new research suggests several ways that feature-highlighting explanations can work around some of the problems that we identify, the disconnect between features in the model and actions in the real world---and the subjective choices necessary to compensate for this---must be understood before these techniques can be usefully implemented.
\end{abstract}

\section{Introduction}
Explanations are increasingly seen as a way to restore the agency that algorithmic decision making takes from its subjects. Advocates believe that explanations can enhance the autonomy of people subject to automated decisions, allow people to navigate the rules that govern their lives, help people recognize when they should contest decisions or object to the decision making process, and facilitate direct oversight and regulation of algorithms~\cite{wachter2018, selbstbarocas2018}.

In this paper, we examine two related approaches to explanation: the counterfactual explanations that have been explored in recent computer science research and which are gaining traction in industry, and the ``principal reason'' approach drawn from U.S. credit laws. We will call these two approaches ``feature-highlighting explanations.'' At a high level, these approaches provide the subject of a decision with a set of factors that ``explain'' the decision. Though they are distinct in operation and motivation, both methods highlight a certain subset of features that are deemed most deserving of the decision subject's attention.

There are at least five reasons for the growing popularity of feature-highlighting explanations. First, this approach appears to allow practitioners to abandon any constraints on model complexity---a constraint often seen as a barrier to improved model performance. Second, it allows businesses to avoid disclosing models in their entirety, thereby protecting trade secrets and businesses' other proprietary interests, while limiting decision subjects' ability to game the model. Third, the approach promises a concrete justification for a decision or precise instructions for achieving a different outcome. Fourth, it allows firms to automate the difficult task of generating explanations for a model's decisions. And finally, it appears to generate explanations that comply with legal requirements both in the United States and Europe.

Generating feature-highlighting explanations is far from straightforward, however, and requires decision makers to make many consequential and subjective choices along the way.
In this paper, we demonstrate that the promised utility of feature-highlighting explanations rests on four key assumptions, easily overlooked, and rarely justified: (1) that a change in feature value clearly maps to an action in the real world; (2) that features can be made commensurate by looking only at the distribution of feature values in the training data; (3) that explanations can be offered without regard to decision making in other areas of people's lives; and (4) that the underlying model is stable over time, monotonic, and limited to binary outcomes.

The paper then explores three tensions at the heart of feature-highlighting explanations. First, while feature-highlighting explanations are designed to respect or enhance the autonomy of decision subjects, the decision maker is put in the position of having to make determinations about what is best for the decision subject; this is paternalism in the name of autonomy. Furthermore, the only way for a decision maker to be sensitive to decision subjects' needs and preferences is to further intrude into their lives, gathering enough information to respect their autonomy, while compromising the autonomy afforded by privacy in the process. Second, partial disclosure puts the decision subject at the mercy of the decision maker. The choice of what to disclose grants a great deal of power to the decision maker. By granting such power to businesses, we invite them to use that power for interests other than those of the decision subject. At best, this leads to beneficent paternalism, at worst, self-dealing. Finally, attempts to overcome some of these challenges by providing decision subjects a larger number and more diverse set of explanations or affording them the opportunity to explore the consequences of specific changes will eventually risk revealing the model altogether. If these techniques fail to protect their intellectual property, firms are unlikely to adopt them.

\section{What are feature-highlighting explanations?}
\label{sec:description}
We define a feature-highlighting explanation as an explanation that seeks to educate the decision subject by pointing to specific features in the model that matter to the individual decision, where each type of feature-highlighting explanation may define ``matter'' differently. The two types we discuss here are counterfactual and ``principal reason'' explanations. Counterfactual explanations in particular have begun to attract the interest of businesses, regulators, and legal scholars, with many converging on the belief that such explanations are the preferred approach to explaining machine learning models and their decisions. Principal-reason explanations are well established in U.S. credit laws, with various businesses having well developed procedures for generating and issuing so-called ``adverse action notices'' (AANs). Both methods aim to produce explanations of a particular decision by highlighting factors deemed useful or important.  There could be other types of feature-highlighting explanations, in principle, but these two are the most developed. This section will describe both approaches, and how they relate to each other.

\subsection{Counterfactual explanations}
\label{sec:counterfactual}

Recent proposals from computer scientists have focused on generating counterfactual explanations for the decisions of a machine learning model~\cite{martens2013explaining,wachter2018,ustun2019actionable,mothilal2019explaining,karimi2019model,hendricks2018generating,grath2018interpretable,ribeiro2016should,lou2012intelligible,dhurandhar2018explanations}. The goal of counterfactual explanations is to provide actionable guidance---to explain how things could have been different and provide a concrete set of steps a consumer might take to achieve a different outcome in the future. Counterfactual explanations are generated by identifying the features that, if minimally changed, would alter the output of the model. 

In particular, an emerging theme in the computer science literature is to frame the search for such features as an optimization problem, seeking to find the ``nearest'' hypothetical point that is classified differently from the point currently in question~\cite{wachter2018,mothilal2019explaining,russell2019efficient,karimi2019model,ustun2019actionable}.
In casting the search for counterfactual explanations as an optimization problem, a key challenge is to define a notion of distance. Different features are rarely directly comparable because they are represented on numerical scales that do not meaningfully map onto one another. We discuss this challenge more in Section~\ref{sec:internal}.

Wachter et al. have also argued that counterfactual explanations could satisfy the explanation requirements of the E.U.'s General Data Protection Regulation (GDPR)~\cite{wachter2018}. Over the last several years, lawyers and legal scholars have debated whether certain provisions of the GDPR create a right to an explanation of algorithmic decisions, and, if it exists, whether and when it requires an explanation of specific decisions or the model~\cite{kaminski2019,selbstpowles2017,brkan2019, wachter2017,edwardsveale2017,casey2019,mendozabygrave2017,malgiericomande2017}. The official interpretation of the Article 29 Working Party---a government body charged with creating official interpretations of European data protection law---has concluded that the GDPR requires, at a minimum, explanations of specific decisions~\cite{A29WP}. Thus, part of the rationale to employ counterfactual explanations is to satisfy the legal requirements of the GDPR.

\subsection{Principal reason explanations}
\label{sec:principal}

The other type of feature-highlighting explanation is what we call a ``principal reason'' explanation. The principal-reason approach has a long history in the United States, where the Fair Credit Reporting Act (FCRA)~\cite{FCRA}, Equal Credit Opportunity Act (ECOA)~\cite{ECOA}, and Regulation B~\cite{RegulationB} require creditors---and others using credit information---to provide consumers with reasons explaining their adverse decisions (e.g., consumers being given a subprime interest rate, denied credit outright, or denied a job based on credit, etc.)~\cite{selbstbarocas2018}. Under ECOA and Regulation B, these decision makers are required to issue AANs to such consumers; under FCRA, consumers are given a list of ``key factors.'' These notices must include a statement of no more than four ``specific reasons'' for the adverse decision~\cite{FCRA, RegulationB}.\footnote{The number four is not a hard limit under Regulation B, as it is under FCRA, but it is observed in practice.} A Sample Form in the Appendix to the regulation offers a non-exhaustive list of acceptable reasons, such as ``income insufficient for amount of credit requested,'' ``unable to verify income,'' ``length of employment,'' ``poor credit performance with us,'' ``bankruptcy,'' and ``no credit file''~\cite[~Appx.~C,~(Sample~Form)]{RegulationB}. Under the regulation, ``no factor that was a \textit{principal reason} for adverse action may be excluded from disclosure, [and t]he creditor must disclose the \textit{actual reasons} for denial''~\cite[~\S~1002.9(b)(2), emphasis added]{RegulationB}.

What counts as a principal reason is not well-defined in either the statutes or regulation. The legislative history of ECOA indicates that consumer education is a primary goal:

\begin{quote}
[R]ejected credit applicants will now be able to learn where and how their credit status is deficient and this information should have a pervasive and valuable educational benefit. Instead of being told only that they do not meet a particular creditor's standards, consumers particularly should benefit from knowing, for example, that the reason for the denial is their short residence in the area, or their recent change of employment, or their already over-extended financial situation~\cite[~p.~4]{SenateReport}
\end{quote}

This would seem to suggest that counterfactual explanations, as currently conceived, would serve the intended purpose of AANs. And indeed, some scholars have suggested as much~\cite{ustun2019actionable}. But this very ambiguity also demonstrates that principal reasons are satisfied by a broader array of possible feature-highlighting explanations. For example, the Official Staff Interpretation to Regulation B, originally published in 1985~\cite{FederalRegister1985}, suggests two ways creditors can generate principal reasons:

\begin{quote}
One method is to identify the factors for which the applicant's score fell furthest below the average score for each of those factors achieved by applicants whose total score was at or slightly above the minimum passing score. Another method is to identify the factors for which the applicant's score fell furthest below the average score for each of those factors achieved by all applicants~\cite[~Supplement I]{RegulationB}.
\end{quote}

Note that neither approach uses the decision boundary as the relevant point of comparison. Instead, they compare the value of applicants' features to the average value of these features for the credit-receiving or general population, in an attempt to surface the dimensions along which the applicant is most deficient.\footnote{Though they are written into the regulation, it is not clear that firms actually use these methods to generate principal reasons.}

Principal-reason explanations lack the precision of counterfactual explanations because they come from the law, not computer science. Legal concepts are rarely defined in mathematically rigorous ways, in no small degree because the ambiguity is meant to be subject to later legal interpretation~\cite{selbst2019fairness}. While we might hope to formally operationalize the concept of principal reasons with these suggested methods, doing so risks imposing a misleading degree of specificity where there is ambiguity.

\subsection{Highlighting subsets of features in the service of autonomy}
For our purposes, counterfactual and principal-reason explanations have one crucial thing in common: neither involves disclosing the model in its entirety. They focus, instead, on highlighting a limited set of features that are most deserving of a decision subject's attention. By design, they do not provide an exhaustive inventory of all the features that a model considers. In practice, learned models can often consider a very large set of features, and an explanation that suggests changes to each of those features would be overwhelming.
As a result, both the law (in the form of principal reasons) and the emerging technical literature (in the form of counterfactual explanations) seek to produce ``sparse'' explanations that present the decision subject with only a small subset of features \cite{wachter2018}.

When opting to use feature-highlighting explanations, there is no natural way to choose between principal-reasons and counterfactual explanations. Yet rarely is the choice to use one method over another discussed explicitly, or, indeed, even recognized as a choice in the first place. These methods produce different explanations and serve fundamentally different goals.

Focusing on features that are furthest from the average value of the features in the credit-receiving or general population casts the problem of identifying principal reasons as one of identifying extreme deficiencies that would seem to rule out the applicant completely, rather than near-misses that applicants might readily address before applying for credit again in the future. While the former may strike us as a less attractive or sensible approach to explanation, there may be good reason to favor an explanation that makes clear the features that were held against an applicant. With the latter approach, while the applicant might receive helpful advice, she might not learn that other features were viewed by the model as crucial marks against her.

Principal-reason explanations treat importance in terms of procedural justice: to respect the autonomy of a decision subject, the decision subject deserves to know which factors dominated the decision~\cite{tyler2006people}. In counterfactual explanations, respect for autonomy means that decision subjects need to know how choices affect outcomes, and thus how they can take actions that will most effectively serve their interests in the future. The former operates more like a justification for a decision---a rationale, with little immediate concern for recourse---whereas the latter serves a more practical purpose---providing explicit guidance for achieving a different decision in the future. Crucially, both styles of explanations can be educational, even if they differ in how easily decision subjects can act on them.

In keeping with these differences, the principal reasons offered by creditors tend to be vaguer (``Income insufficient for amount of credit requested''), while counterfactual explanations aim for precision (``Had you earned \$5,000 more, your request for credit would have been approved''). In many cases, principal-reason explanations do not even disclose the magnitude, let alone the direction, of change that would be necessary to achieve a different outcome, while such details are inherent to counterfactual explanations. In fact, the official interpretation of the regulation that requires creditors to issue AANs notes that creditors "need not describe how or why a factor adversely affected an applicant. For example, the notice may say ``length of residence'' rather than ``too short a period of residence''~\cite[~\S~1002.9(b)(2)~(Official~Interpretation)]{RegulationB}

\section{Feature-highlighting explanations in practice}
\label{sec:analysis}
There are several hidden assumptions behind the belief that feature-highlighting explanations will be useful for decision subjects. In this section, we identify four such assumptions, explain why they might not be valid, and explore the consequences of that realization.

\subsection{Features do not clearly map to actions}
\label{sec:actions}
Feature-highlighting explanations often assume a clear and direct translation from suggested changes in feature values to actions in the real world. In many cases, this is a reasonable assumption: instructing someone to reduce their total lines of credit maps onto the obvious action of cancelling a credit card or fully repaying---and thus dispensing with---a loan. In most of the contexts in which existing scholarship considers the challenge of explaining the decisions of a machine learning model, there is a clear correspondence between the feature values that one is told to change and the actions that one would take to achieve those changes.
And yet, in many cases, we are only able to perform this mapping because we have relevant domain knowledge and an implicit causal model in mind that relates specific actions in the real world to predictable changes in feature values.

\begin{figure}[ht]
    \centering
    \includegraphics[width=.4\textwidth]{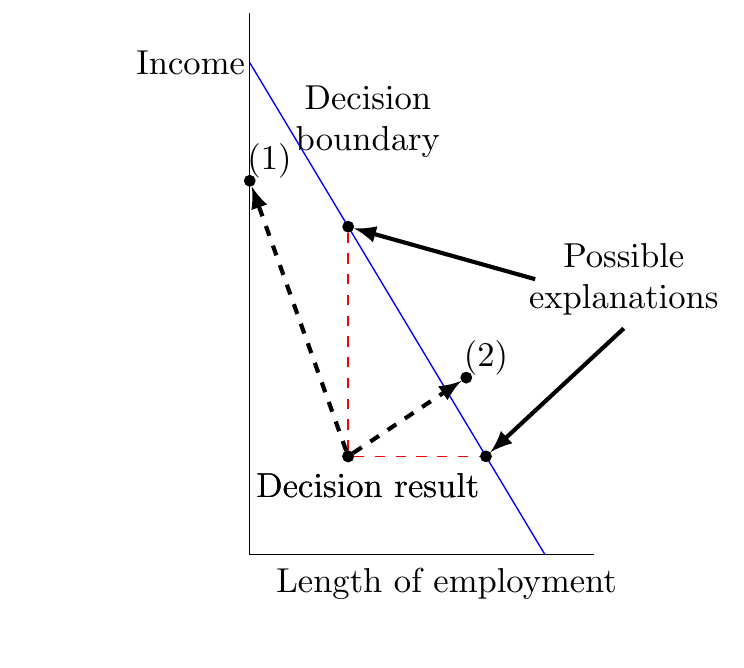}
    \caption{A decision based on two features---income and length of employment---will be explained by reference to one of the features, either the shortest or longest distance from the boundary. But the explanations do not map to the decision subject's possible actions that can affect them. Point (1) represents getting a higher-paying job, and point (2) represents waiting for a raise.
        }
    \label{fig:graph}
\end{figure}

Even when the highlighted feature seems to refer to something rather concrete, the actions a decision subject can take to affect those features may not line up with the features themselves. For example, a recommendation that someone increase his income can lead a person to take one of several actions: he can seek a new job, ask for a raise, or take on more hours. As Figure~\ref{fig:graph} illustrates, these actions are not as simple as ``increase income'' or ``increase length of employment.''

Individuals cannot in general instantaneously change the features suggested by an explanation.
Some features, like length of employment, are inherently time-dependent.
Other changes may take varying degrees of time to implement.
Suppose, for instance, the decision subject could obtain credit either by increasing their income by \$5,000 or by increasing the length of their employment by 6 months. If it would take 6 months of especially grueling work to secure a raise, it would seem unnecessary for him to do so instead of simply waiting 6 months to qualify. On the other hand, if the subject needs credit immediately and is able to change to a higher-paying job right away, then we might view income as the right feature to provide the subject in an explanation. Because the relationship between features and actions is time-dependent, explanations that do not consider temporal aspects will fail to highlight the ``right'' features.

But these examples still assume a relatively direct relationship. To act on explanations that instruct us to change certain feature values, we need to know what causes features to change value in the real world. This might not be obvious; we may struggle to identify the actions that would cause a feature value to change---or change in a predictable way.

Recent work on ``actionable'' explanations has focused on avoiding explanations that tell people to make changes that are impossible, placing the burden on decision makers to give advice that is sensitive to the actual steps that decision subjects would need to take to achieve the change in feature values~\cite{ustun2019actionable, mothilal2019explaining}. Avoiding these potential explanations is a matter of identifying the lack of any possible causal mechanism in the real world that would have the necessary effect on the value of some feature.

``Gaming'' is just a special case of this disconnect between feature changes and actions~\cite{bambauer2018algorithm}. When a decision maker instructs someone to change certain features, the decision maker will often assume that the person will take a specific desirable sequence of actions because that is the causal mechanism that the decision maker has in mind. But there are often many other ways to change feature values that don't require taking these steps~\cite{kleinberg2019classifiers}.

Highlighting certain features as those that need to change to obtain a different decision also implicitly relies on the belief that everything else can be held constant while making these changes. In reality, actions may affect multiple features simultaneously. Changes in the value of one feature may also affect the value of another feature, if the two features interact. As Figure~\ref{fig:graph} demonstrates, whether one increases his income by finding a higher-paying job or waiting for his performance review to get his raise, the action will affect both income \textit{and} length of employment, a separate feature in the model. In the case of a job change, length of employment will be negatively affected. Thus, increasing income may not be enough to get credit, which is why point (1) is on the left of the decision boundary. In the case of waiting for a raise, a smaller increase in income might be needed than the explanation would say, because length of employment increases at the same time. This is why point (2) is on the right side of the decision boundary, despite not increasing income as much as the explanation suggests. When considered in the context of feasible real-world actions, it becomes clear that features may not be \textit{independent}: changing one feature may impact others.

Insisting that explanations exhibit sensitivity to these constraints is analogous to insisting that explanations consider the causal mechanisms that allow decision subjects to alter the value of specific features. Indeed, the only way to ensure that the recommended change is even possible, to prevent gaming, and to account for dependencies between features is to model the outcome of interest using features that directly figure into the causal mechanism. The idea that we can identify all such constraints in advance assumes that the actions necessary to change specific feature values will always be self-evident.

\subsection{Features cannot be made commensurate by looking only at the distribution of the training data}
\label{sec:internal}

All feature-highlighting explanations rely on some notion of a distance between the observed values for various features and some reference point, whether the the decision boundary or the average value in the population. Relying on distance requires normalizing features, because there needs to be a shared scale between features in order to meaningfully compare them. For example, as discussed in Section~\ref{sec:counterfactual}, an increase in length of employment is not naturally commensurate with an increase in salary. Normalization attempts to capture the fact that salaries may vary on the order of thousands or tens of thousands of dollars, but length of employment (in years, say) varies at a numerically much smaller scale.

Several statistical techniques exist to address this problem, scaling features so as to make them seemingly comparable, and different explanation techniques use different approaches. Following Wachter et al., the literature on counterfactual explanations has mostly converged on a heuristic that finds the Median Absolute Deviation (MAD) under an L1 distance norm~\cite{russell2019efficient,mothilal2019explaining,grath2018interpretable}. Meanwhile, it is entirely unclear what methods principal-reason explanations use---the regulations do not specify, and it is never discussed in practice---but the nature of a distance metric requires that \textit{something} be used. 
Normalization techniques are typically based entirely on the distribution of the data, not some external point of reference.

When examined from the perspective of a decision subject who must take some action in response to these explanations, normalization based simply on the distribution of data is somewhat arbitrary. One decision maker might scale the axes such that increasing income by \$5,000 annually is equivalent to an additional year on the job. A competing lender, using different training data, could conclude that \$10,000 of income corresponds to one year of work. These lenders might therefore produce different explanations depending on the scaling of attributes. Without an external point of reference to ground these scales, the meaning of the relative difference in feature values is unclear. 

Relying on the distribution of the training data to normalize features is an example of what Selbst et al. have called the ``framing trap,'' where an attempt to address concerns with accountability targets properties of the model alone, ignoring crucial interactions with the outside world~\cite{selbst2019fairness}. What counts as seemingly equivalent features according to the distribution of the training data is not necessarily what the decision subject would view as equivalent. Making features commensurate in this way fails to consider the particular circumstances faced by any given decision subject.

Because we are concerned with decision subject's menu of options, the most sensible external referent would be something akin to the cost of making the required change, where cost can imply dollars spent, effort, or time. For counterfactual explanations, those features that involve little \textit{cost} to change, even if they involve considerable change along a normalized numeric scale, may be far more useful to highlight than those that would be costlier to change.
If, instead, the preferred approach involves generating principal reasons, features that are costly or impossible to change may be precisely the ones that should be highlighted. Given its focus on identifying the features easiest to change, counterfactual explanations might conceal that a decision largely hinged on immutable characteristics like race. Worse, such explanations might mislead the subjects of these decisions into believing that their fate was determined by factors under their control. This is particularly worrying, considering that the field’s turn to explanations has in large part been a response to concerns about discrimination or fairness and the statute that mandates AANs is itself an anti-discrimination law. Knowing that one's application for credit has been rejected due to characteristics outside one's control might be paramount if the goal is to ensure procedural justice or to reveal when to contest the decision~\cite{mendozabygrave2017,wachter2018,brkan2019}.

Thinking about changes in terms of their real-world cost therefore helps to translate numerical changes in feature values to real-world actions, whether we want to point out either what is easiest or most difficult to change. Some recent work seeks to account for the cost of actually manipulating features in practice by assuming domain knowledge or soliciting user input~\cite{ustun2019actionable,grath2018interpretable,mothilal2019explaining}, but decision subjects may be unable to articulate all of the relevant real-world constraints that would affect the utility of an explanation.

Worse yet, the cost of making certain changes will not be consistent across different people. Changes that might be rather inexpensive for one person to make might be costly for another person to make~\cite{rudin2019stop}. Thus, when we use explanations to identify the easiest or most difficult features for someone to change to achieve a different decision from a model, the explanation must be sensitive not only to how these changes involve different costs, but how these costs vary across the population. Different subsets of features may be appropriate for different people with different life circumstances. This complication cuts against the very desirability of these explanations: the idea that we can automatically determine what is easiest or hardest to change.

\subsection{Features may be relevant to decision making in multiple domains}

Feature-highlighting explanations may interact with facts about a person's life that are invisible to the model. In particular, the supposition of a counterfactual explanation is that it is offering advice about the kinds of changes that it would be rational for a person to make to achieve better results in future decisions. Some commentators and scholars have cautioned that explanations should never encourage people to take actions that are irrational or harmful~\cite{oreilly2017, eubanks2018}. What they mean more specifically is that there may be some recommendations that are indeed rational if the only goal is to obtain a positive decision from the model, but irrational with respect to other goals in a person's life.

A common-sense example for this proposition is that an explanation should never recommend that a person seek to make less money~\cite[e.g.~][]{lipton2017}. While we believe it unrealistic that an actual credit model would ever (be allowed to) learn such a relationship, the example still holds value. It is self-evident that no one would want to make less money, even if doing so would improve their access to credit. Or consider an example that reverses this dependency: a person contemplating applying to a new job for its superior health insurance is unlikely to remain at his current job because an explanation for a failed credit application told him to increase the value of his length-of-employment feature. In this case, acting on the recommendation would impose an opportunity cost on the consumer by forcing him to forgo benefits in other domains. When other aspects of one's life depend on some of the same features, explanations for how to get the desired outcome in one aspect of your life may conflict with those in another.

We can reason about this the other way around as well. From the point of view of a counterfactual explanation, an applicant might be best off trying to change a number of other features \textit{besides} income. Yet, from the perspective of the applicant, increasing her income might have ancillary benefits in other parts of her life that make this change more attractive---and indeed rational---than those suggested by the explanation. Increasing her income would grant her improved access not only to credit, but to improved quality of life, generally.

In the first case, a change in feature might benefit the decision subject in one domain, while hurting her in others. In the second case, a change in a feature might benefit the decision subject in multiple domains, not just one. These spillovers---both negative and positive---complicate the process of determining which features would be most useful to highlight in an explanation. Ideally, feature-highlighting explanations would allow decision subjects to avoid negative spillovers and identify opportunities for positive spillover. But a decision maker will lack information about the many other goals that a person might have in her life and the features that are relevant in those domains. 

This also highlights an additional risk: due to other life goals, decision subjects may change \textit{undisclosed} features unless otherwise instructed. For example, if a counterfactual explanation tells someone to increase her income and lower her debt, but fails to mention that she should not reduce her length of employment, she may have no idea that she should avoid any career change while attempting to address these other issues, stumbling accidentally into point (1) in Figure~\ref{fig:graph}. Indeed, she might not even know that length of employment figured into the credit decision in the first place. Thus, by failing to disclose what a decision subject must \textit{not} change, an explanation may lead her to take an ultimately unsuccessful action.

\subsection{Models may not have certain properties: stability, monotonicity, and binary outcomes}
\label{sec:binary}

Those advocating in favor of feature-highlighting explanations tend to assume that the underlying models have certain properties: stability, monotonicity, and binary outcomes.

Credit scoring models are routinely retrained to react to changes in the overall environment and to changes in borrowers' behavioral patterns. Perhaps there is another recession. Or perhaps borrowers change their behavior en masse based on the very explanations that lenders have offered in the past, shifting the data distribution over time~\cite{liu2018delayed}. Any of these changes would necessitate model retraining by the lender. Wachter et al. have thus argued that the law should treat a counterfactual explanation as a promise rather than just an explanation~\cite{wachter2018}. They argue that if a rejected applicant makes the recommended changes, the promise should be honored and credit granted, irrespective of the changes to the model that have occurred in the meantime. Whether this is the right approach or not, it is a recognition that without such a guarantee, counterfactual explanations might not serve their purpose when one considers the time it takes to make the suggested changes.

Feature-highlighting explanations can also be misleading if the model has not been subject to monotonicity constraints, which guarantee that as the value of the features move in the recommended direction, the decision subject's chances of success consistently improves. Without monotinicity constraints, a model might learn complex and even counter-intuitive relationships between certain features and an outcome of interest. For example, a model might learn that people who have spent two to four years at their current job are good candidates for credit, while those who have stayed five or more are not. Likewise, carrying more debt might render applicants less attractive, until they start earning more income, at which point additional debt might make them more attractive. 

Decision subjects will not necessarily be able to alter the value of these features through some sudden step change. Instead, they may have to make incremental changes in the direction of the specified value. And despite their best efforts, decision subjects might struggle to hit the specified feature value; their efforts could move the value of these features in the right direction, but ultimately fail to get the decision subject all the way there. Similarly, decision subjects might lack precise control over the value of a feature, making it difficult to avoid overshooting the mark when they take some action. Unless the model exhibits monotonicity with respect to the highlighted features, the decision subject might find herself in a \textit{worse} position as she moves toward the specified value or if she exceeds it. 

Similarly, principal-reasons explanations can lose their utility without monotonicity. In some cases, highlighted features like ``length of employment'' do not suggest an obvious direction for improvement, and indeed, models may not be monotone in such features. In such cases, a decision subject will not know whether an adverse decision was made because the highlighted feature was too high or too low---simply that it was particularly far from what was expected.

Finally, in the computer science literature, model explanations often assume that outcomes are binary: did the applicant receive a loan or not? In reality, the creditor decides not only whether or not a loan is given, but also the loan's interest rate. How should feature-highlighting explanations account for this? Does the decision maker choose a specific target interest rate when providing an explanation? What if the applicant is only interested in a loan below a certain rate?

Consider a financially responsible borrower who will only accept a loan at a sufficiently low interest rate. If she is told via a counterfactual explanation that she could qualify for a high-interest loan by increasing her income without reducing her debt, she learns nothing about how to qualify for a low-interest loan; what it would take to obtain a low-interest loan might be very different than what it would take to secure one with a high interest rate.
There may be no way to extrapolate a strategy for obtaining a low-interest loan from the counterfactual explanation that gets her to a high-interest loan. Indeed, she may not even know that the counterfactual explanation that tells her how to get a loan is specific to a high-interest loan, instead seeing the interest rate on offer as the only option, and concluding that she cannot get a better rate.

\section{Unavoidable Tensions}
We have argued so far that the need to disclose a limited subset of features infuses feature-highlighting explanations with subjective choices and creates a number of challenges that makes their promise harder to realize in practice than advocates of such techniques would have us believe. They also present a number of unavoidable normative tensions. Decision makers start with a great deal of power over decision subjects, and the purpose of explanations---and the legal requirements for them---is to restore some degree of power to the decision subjects. Yet the fact that decision makers must, by necessity, withhold information creates three unavoidable tensions. First, in order to generate genuinely helpful explanations, decision makers must be both paternalistic and privacy-invasive. That is, they must interfere with decision subjects' autonomy to offer some back to them. Second, while designed to restore power to decision subjects, partial explanations grant a new kind of power to the decision maker, to use for good or to abuse as desired. Finally, the additional transparency that might help overcome some of the problems with feature-highlighting explanations render these techniques anathema to the decision makers that we would want to use them.

\subsection{The autonomy paradox}

Feature-highlighting explanations are motivated by either the desire to make recommendations to decision subjects or to justify the model's decision. Both these motivations originate from a concern with the autonomy of decision subjects. Recommendations appeal to an instrumental vision of autonomy, where information enables action. Justification, however, is more focused on a moral conception; the information is due because the subject deserves to know. Both of these motivations are complicated by the need to withhold some information.

Ironically, respecting a decision subject's autonomy will require making assumptions about which information will be valuable to a given decision subject. The decision maker will not know how features correspond to actions in the real world, and thus which features a decision subject could most readily change.
The decision maker does not know how a change in the measured feature, or the action required to make such a change, affects other aspects of a person's life positively or negatively. And assuming the decision maker could achieve a general sense of these facts, the decision maker further does not know how they vary from person to person. All this means that the choice of features to disclose can have unintended effects for decision subjects, which would have been avoided with a different disclosure. Given the informational position of the decision maker, there is simply no way to fully realize its commitment to respecting a decision subject's autonomy.

One might suggest that these problems can be solved with even more data. If explaining a credit decision, the decision maker's choice of which features to highlight might be affected by information about a person's health, family situation, or future educational plans. And a decision subject's willingness to look for a new job will be influenced by whether they are sick, have a new baby or aging relative to care for, or are saving to go back to school. This information can directly or indirectly be mined from other sources in the world, such as social media data. If a decision maker understands other aspects of a person's life that may interact with the decision, then it might be able to offer explanations that are appropriate and tailored, or might be able to focus on features that are relevant to decisions in multiple contexts. So perhaps the answer is to collect it all.

Unfortunately, allowing a decision maker, such as a lender, to collect and connect every bit of information about a person's life is not really a solution. Rather, it is a privacy disaster~\cite[e.g.~][]{solove2006, cohen2012configuring, nissenbaum2009}, and because privacy is a fundamental aspect of autonomy~\cite{cohen2012privacy, reiman1976}, this leads to an autonomy paradox. The problem appears most clearly through the lens of Helen Nissenbaum's theory of contextual integrity~\cite{nissenbaum2009}. Contextual integrity argues that a privacy violation occurs where information flows between actors in social context in ways that violate the informational norms relative to that context. So in one sense, this solution---allowing lenders in financial context to access social media information---is definitionally problematic. A primary concern of contextual integrity is for social contexts to keep operating as they should. If creditors have access to social media data, the worry is that they will make credit determinations based on public friendships, for example, and as a result people will have incentives to change or hide their social relationships in order to get better credit~\cite{packinlevaretz2016}. This will harm the social contexts in which friendships flourish for the sake of credit. Ironically, then, while in Section~\ref{sec:analysis} we argued that explanations would not be useful unless the decision maker understood facts about people's lives beyond those considered in the model, contextual integrity would suggest that the fact of decision makers knowing this information is itself harmful to autonomy. 

But imagine a decision subject who, facing an adverse credit decision, is shown the complete model and finds it overwhelming. Such a person may instead prefer the counterfactual explanation, even if it requires the decision subject to disclose all the information necessary for the decision maker to offer an appropriately tailored set of instructions. This exact scenario motivates much of the work on counterfactual explanations, so we should not discount that a more informed explanation can still be autonomy-enhancing on balance. In a sense, this tension is reflective of a common concern in discussions of autonomy: When can giving up information and agency be autonomy-enhancing? For example, if a person hires an attorney, she outsources some important decisions, gives up very private information, and often gets answers back that she cannot understand. But it is doubtful that anyone would consider hiring a lawyer to be a loss of autonomy. At the same time, we would immediately recognize that furnishing lenders with detailed information and relinquishing control over decision making is not an obvious mechanism for enhancing one’s autonomy. Notably, there are no requirements that they act in your best interest, while such fiduciary obligations do apply to lawyers. This is a difficult tension to resolve, and may depend on the relative power of and constraints upon the decision maker, rather than the quality of the explanation.

\subsection{The burden and power to choose}
\label{sec:power}

One of the reasons feature-highlighting explanations are so appealing is that they appear to offer complete automation: whenever a decision is made, an explanation can be provided without any further human intervention. But this veneer of mechanization belies the fact that such explanations cannot be completely formulaic. They require decisions about what to disclose and assumptions about the real world.

The need for partial disclosure grants new power to the decision maker. Of course, a decision maker---by virtue of being one---has always had power over the decision subject. But by attempting to return power to the decision subject via an explanation that, for her own sake, cannot be a complete explanation, we grant a new form of largely unanticipated power to the decision maker. Furthermore, the requirement to make certain assumptions about the real world also grants power to the decision maker. Whenever there is ambiguity in the individual's preferences, the decision maker has the power to resolve the ambiguity however it sees fit. This leaves the decision maker with significant room to maneuver, the choice of when and where to further investigate, and more degrees of freedom to make choices that promote their own welfare than we might realize.

This new power can be used for good or ill. Consider, for example, a decision maker providing a counterfactual explanation for why an individual did not qualify for a loan. As discussed in Section~\ref{sec:binary}, this decision (and therefore explanation) is not simply binary---in its explanation, the decision maker must give the decision subject a counterfactual that would result in a \textit{specific} interest rate. At best, it might allow the subject to choose their target interest rate. Alternatively, it might---somewhat paternalistically---choose the interest rate that it believes is ``right'' for this subject. More insidiously, it might choose the interest rate that is likely to maximize its profit. Ultimately, the point is that in the absence of standards or robust avenues for user input, the decision maker is left with the power to make this decision on its own.

This power is not simply limited to the choice of outcome. As we have argued, many aspects of explanations are unspecified by the law or by technical proposals, including what factors can be included in an explanation, what the relative costs of various features are, and how to account for real-world dependencies between them. The key point here is that left to their own devices, decision makers are afforded a remarkable degree of power to pursue their own welfare through these choices.

\subsection{Too much transparency}

Decision makers might seek out different ways to address the difficulty of taking decision subjects' real-world circumstances into account when generating feature-highlighting explanations. A number of recent papers propose presenting the user with a diverse set of counterfactual explanations~\cite{mothilal2019explaining,russell2019efficient,wachter2018}, allowing the decision subject to choose among several possible ways to achieve a favorable outcome. This approach accepts that decision makers may lack the capacity to ever fully account for the unique constraints and preferences of decision subjects, instead providing a wide range of possible paths to success from which the decision subjects can choose. Doing so allows the decision subject to rely on knowledge of her own particular circumstances in selecting among these. 

Others have advocated in favor of interactive tools that allow decision subjects to explore the effect of making changes to certain features~\cite{citronpasquale2014,hildebrandt2006}. Industry has even implemented some such tools.\footnote{See, e.g., Credit Karma's Credit Score Simulator: \url{https://www.creditkarma.com/tools/credit-score-simulator/}} This approach gives decision subjects greater freedom to explore the space, using a deep understanding of their own constraints and preferences to investigate the effect of certain adjustments.

Still other work adopts an entirely different approach, focusing instead on finding ways for the decision maker to learn more about decision subjects. In particular, there have been recent proposals to devise mechanisms for soliciting input from decision subjects, allowing them to communicate whether they find certain counterfactuals helpful, whether changes to certain features are out of the question or less desirable, and what other preferences they might hold~\cite {mothilal2019explaining}. 

These approaches could also work in concert, seeding decisions subjects with an initial set of diverse explanations that could serve as starting points for interactive exploration. In theory, this would have the benefit of helping to ensure that decision subjects do not fail to explore the space sufficiently, concluding their investigation after only making a small number of adjustments from one initial starting point.

Unfortunately, each of these approaches runs the risk of revealing a sufficient amount of information about the underlying model to reconstruct it~\cite{tramer2016stealing}. As a result, while these approaches may be the most promising to overcome certain difficulties, they create difficulties of their own. Firms concerned with intellectual property and gaming are unlikely to afford decision subjects extensive freedom to explore.

\section{Conclusion}
Feature-highlighting explanations have been embraced as a way to help decision makers avoid a number of difficult trade-offs, granting firms the capacity to provide meaningful and useful explanations of machine-learned models without having to compromise on model performance, while also respecting concerns with trade secrecy, gaming, and legal compliance. Advocates have championed this style of explanation as an elegant way to honor and enhance decision subjects' autonomy even as machine learning models grow in complexity and ubiquity.

Yet as we have shown, these explanations lack a connection to the real-world actions required to change features. They fail to consider the cost of these actions, decision subjects' preferences, and the effects of the necessary actions on other parts of decision subjects' lives. Worse, attempts to correct these deficiencies undermine the very goals of explanation by violating decision subjects’ autonomy in the name of enhancing it and granting more power to decision makers when trying to return it to decision subjects.

So what can be done? How can feature-highlighting explanations be useful, while protecting the autonomy of decision subjects? Much more work is needed to address the issues we have raised here, but we see three concrete avenues worth exploring.

First, at an absolute minimum, given the power that these explanations grant to decision makers, they should disclose the method by which they generate explanations. Additionally, legal requirements for explanation and AANs should be amended to require this. Without understanding the method of explanation, decision subjects have no hope of understanding how to effectively realize their goals.

Second, to address the autonomy paradox, fiduciary obligations are worth exploring. Fiduciary obligations are legal requirements that constrain certain people or entities who hold a position of trust to act in the best interests of their beneficiaries. Common examples or fiduciary-beneficiary pairs include attorneys and clients, corporate officers and shareholders, and in some cases, financial advisors and advisees. Over the last several years, legal scholars have debated whether an “information fiduciaries” framework would work well as a scheme to regulate the data economy, generally~\cite[e.g.][]{balkin2015information, khan2019skeptical}. While these scholars are not primarily writing about feature-highlighting explanations, the case for fiduciary obligations seems especially good here, as we observe a similar situation to that of any position of trust: (1) there may be no way to explain the decisions of a machine learning model sufficiently well that a person can devise a rational course of action for herself and (2) the institution charged with explaining its decisions might otherwise have their own interests in suggesting a specific course of action. As a policy matter, a fiduciary obligation on the institution would require that its explanations are aligned with the subject’s best interests, and may help resolve some of tensions created by the autonomy paradox and power to selectively disclose features.

Third, we need to understand what actions people actually take when confronted with feature-highlighting explanations---and which disclosures help people act most effectively. Empirical research is essential to answer these questions. One obvious place to start such work is with longitudinal data documenting the successful paths that previous decision subjects have taken to receive positive outcomes when starting from various circumstances. Another is to engage directly with decision subjects to develop a richer account of their everyday strategies for responding to models and explanations of their decisions, as Malte Ziewitz and Ranjit Singh have done over the past few years.\footnote{https://zwtz.org/restoring-credit/} This approach rests on the idea that explanations should focus on communicating what had worked well for other people under seemingly similar conditions.

These proposals will only address some of the issues with feature-highlighting explanations raised here. In work concurrent with our own, Venkatasubramanian and Alfano raise similar concerns and offer some potential solutions~
\cite{venkatasubramanian2020philophical}. Going forward, there is still more work to do, in computer science, social science, and policy, if we want to understand when and where feature-highlighting explanations can be useful to decision makers and decision subjects alike.

\paragraph{Acknowledgments}
We thank Kiel Brennan-Marquez, Albert Chang, Madeleine Elish, Sorelle Friedler, and Jon Kleinberg for their feedback and suggestions. Funded in part by the NSF (IIS-1633400) and a Microsoft Research PhD Fellowship.

\bibliographystyle{plain}
\bibliography{refs}

\end{document}